\documentstyle[12pt]{article}
\begin{document}

%%%%%%%%%%%%%%%%%%%%%%%%%%%%%%%%%%%%%%%%%%%%%%%%%%%%%%%%%%%%%%
\catcode`@=11
% Redefine caption to put text and formulas in smaller font
\long\def\@caption#1[#2]#3{\par\addcontentsline{\csname
  ext@#1\endcsname}{#1}{\protect\numberline{\csname
  the#1\endcsname}{\ignorespaces #2}}\begingroup
    \small
    \@parboxrestore
    \@makecaption{\csname fnum@#1\endcsname}{\ignorespaces #3}\par
  \endgroup}
\catcode`@=12
%%%%%%%%%%%%%%%%%%%%%%%%%%%%%%%%%%%%%%%%%%%%%%%%%%%%%%%%%%%%
\newcommand{\newc}{\newcommand}
\newc{\gsim}{\lower.7ex\hbox{$\;\stackrel{\textstyle>}{\sim}\;$}}
\newc{\lsim}{\lower.7ex\hbox{$\;\stackrel{\textstyle<}{\sim}\;$}}
\newc{\gev}{\,{\rm GeV}}
\newc{\mev}{\,{\rm MeV}}
\newc{\ev}{\,{\rm eV}}
\newc{\kev}{\,{\rm keV}}
\newc{\tev}{\,{\rm TeV}}
\def\tr{\mathop{\rm tr}}
\def\Tr{\mathop{\rm Tr}}
\def\Im{\mathop{\rm Im}}
\def\Re{\mathop{\rm Re}}
\def\bR{\mathop{\bf R}}
\def\bC{\mathop{\bf C}}
\def\lie{\mathop{\hbox{\it\$}}} %pound sterling
\newc{\sw}{s_W}
\newc{\cw}{c_W}
\newc{\swsq}{s^2_W}
\newc{\cwsq}{c^2_W}
\newc{\mgrav}{m_{3/2}}
\newc{\mz}{M_Z}
\newc{\mpl}{M_*}
\def\ux{U(1)$_X$}
%\renewcommand{\phi}{\varphi}
%
%%%%%%%%%%%%%%%%%% Reference Defs %%%%%%%%%%%%%%%%%%
%
\def\NPB#1#2#3{Nucl. Phys. {\bf B#1} (19#2) #3}
\def\PLB#1#2#3{Phys. Lett. {\bf B#1} (19#2) #3}
\def\PLBold#1#2#3{Phys. Lett. {\bf#1B} (19#2) #3}
\def\PRD#1#2#3{Phys. Rev. {\bf D#1} (19#2) #3}
\def\PRL#1#2#3{Phys. Rev. Lett. {\bf#1} (19#2) #3}
\def\PRT#1#2#3{Phys. Rep. {\bf#1} (19#2) #3}
\def\ARAA#1#2#3{Ann. Rev. Astron. Astrophys. {\bf#1} (19#2) #3}
\def\ARNP#1#2#3{Ann. Rev. Nucl. Part. Sci. {\bf#1} (19#2) #3}
\def\MPL#1#2#3{Mod. Phys. Lett. {\bf #1} (19#2) #3}
\def\ZPC#1#2#3{Zeit. f\"ur Physik {\bf C#1} (19#2) #3}
\def\APJ#1#2#3{Ap. J. {\bf #1} (19#2) #3}
\def\AP#1#2#3{{Ann. Phys. } {\bf #1} (19#2) #3}
\def\RMP#1#2#3{{Rev. Mod. Phys. } {\bf #1} (19#2) #3}
\def\CMP#1#2#3{{Comm. Math. Phys. } {\bf #1} (19#2) #3}
\relax
%
%
%%%%%%%%%%%%%%%%%%%%%%% latex eqn abrev's %%%%%%%%%%%%%%%%%%%%%%%%%%%%
%
\def\beq{\begin{equation}}
\def\eeq{\end{equation}}
\def\bea{\begin{eqnarray}}
\def\eea{\end{eqnarray}}
%
%%%%%%%%%%%%%%%%%%% special features for eqns %%%%%%%%%%%%%%%%
%
%     this boxes an equation
%
\def\boxeqn#1{\vcenter{\vbox{\hrule\hbox{\vrule\kern3pt\vbox{\kern3pt
\hbox{${\displaystyle #1}$}\kern3pt}\kern3pt\vrule}\hrule}}}
%
%     this draws a little box (end of proof symbol)
%     e.g. \qed{.1}{.1}
%
\def\qed#1#2{\vcenter{\hrule \hbox{\vrule height#2in
\kern#1in \vrule} \hrule}}
\def\half{{\textstyle{1\over2}}} %%small half in a displayed eqn
%\def\frac#1#2{{\textstyle{#1\over #2}}} %%small fraction in a displayed eqn
%
%
%%%%%%%%%%%%%%%%%%%%%%% common abrev's %%%%%%%%%%%%%%%%%
%
%
\newc{\ie}{{\it i.e.}}          \newc{\etal}{{\it et al.}}
\newc{\eg}{{\it e.g.}}          \newc{\etc}{{\it etc.}}
\newc{\cf}{{\it c.f.}}
%
%
%%%%%%%%%%%%%%%%%%%%%%%% curly letters %%%%%%%%%%%%%%%%%%%
%
%
\def\CAG{{\cal A/\cal G}}
\def\CA{{\cal A}} \def\CB{{\cal B}} \def\CC{{\cal C}} \def\CD{{\cal D}}
\def\CE{{\cal E}} \def\CF{{\cal F}} \def\CG{{\cal G}} \def\CH{{\cal H}}
\def\CI{{\cal I}} \def\CJ{{\cal J}} \def\CK{{\cal K}} \def\CL{{\cal L}}
\def\CM{{\cal M}} \def\CN{{\cal N}} \def\CO{{\cal O}} \def\CP{{\cal P}}
\def\CQ{{\cal Q}} \def\CR{{\cal R}} \def\CS{{\cal S}} \def\CT{{\cal T}}
\def\CU{{\cal U}} \def\CV{{\cal V}} \def\CW{{\cal W}} \def\CX{{\cal X}}
\def\CY{{\cal Y}} \def\CZ{{\cal Z}}
%
%
%
%%%%%%%%%%%%%%%%%%% derivatives %%%%%%%%%%%%%%%%%%%%%%%%%%%%%
%
%
\def\grad#1{\,\nabla\!_{{#1}}\,}
\def\gradgrad#1#2{\,\nabla\!_{{#1}}\nabla\!_{{#2}}\,}
\def\partder#1#2{{\partial #1\over\partial #2}}
\def\secder#1#2#3{{\partial^2 #1\over\partial #2 \partial #3}}
%
%
%
%%%%%%%%%%%%%%%%%%%%% relations %%%%%%%%%%%%%%%%%%%%%%%%%%%%%
%
%
\def\ltap{\ \raise.3ex\hbox{$<$\kern-.75em\lower1ex\hbox{$\sim$}}\ }
\def\gtap{\ \raise.3ex\hbox{$>$\kern-.75em\lower1ex\hbox{$\sim$}}\ }
\def\gl{\ \raise.5ex\hbox{$>$}\kern-.8em\lower.5ex\hbox{$<$}\ }
\def\roughly#1{\raise.3ex\hbox{$#1$\kern-.75em\lower1ex\hbox{$\sim$}}}
%
%
%%%%%%%%%%%%%%%%%%%% slashed symbols %%%%%%%%%%%%%%%%%%%%%
%
%
\def\slash#1{\rlap{$#1$}/} % slashes a character
\def\dsl{\,\raise.15ex\hbox{/}\mkern-13.5mu D} %this one can be subscripted
\def\delsl{\raise.15ex\hbox{/}\kern-.57em\partial}
\def\Ksl{\hbox{/\kern-.6000em\rm K}}
\def\Asl{\hbox{/\kern-.6500em \rm A}}
\def\Dsl{\hbox{/\kern-.6000em\rm D}} %roman D
\def\Qsl{\hbox{/\kern-.6000em\rm Q}}
\def\gradsl{\hbox{/\kern-.6500em$\nabla$}}
%
%%%%%%%%%%%%%%%%%%% greek letters %%%%%%%%%%%%%%%%%%%%
%
\let\al=\alpha
\let\be=\beta
\let\ga=\gamma
\let\Ga=\Gamma
\let\de=\delta
\let\De=\Delta
\let\ep=\varepsilon
\let\ze=\zeta
\let\ka=\kappa
\let\la=\lambda
\let\La=\Lambda
\let\del=\nabla
\let\si=\sigma
\let\Si=\Sigma
\let\th=\theta
\let\Up=\Upsilon
\let\om=\omega
\let\Om=\Omega
\def\ph{\varphi}
%
%%%%%%%%%%%%%%%%% BOLD greek letters %%%%%%%%%%%%%%%%%%%%%%%%%%%%%%
%
% Style-sensitive Poor-Man's-Bold command, produces bold greek letters.
% Usage $ ... \pmb\gamma ... $
% Adapted from TeXbook p386 (\pmb) and p360 (\mathpallette)
%
\newdimen\pmboffset
\pmboffset 0.022em
\def\oldpmb#1{\setbox0=\hbox{#1}%
 \copy0\kern-\wd0
 \kern\pmboffset\raise 1.732\pmboffset\copy0\kern-\wd0
 \kern\pmboffset\box0}
\def\pmb#1{\mathchoice{\oldpmb{$\displaystyle#1$}}{\oldpmb{$\textstyle#1$}}
        {\oldpmb{$\scriptstyle#1$}}{\oldpmb{$\scriptscriptstyle#1$}}}
%
%
%
%%%%%%%%%%%%%%%%%%% various symbol abbreviations, vev's etc %%%%%%%%%%%
%
%
\def\bar#1{\overline{#1}}
\def\vev#1{\left\langle #1 \right\rangle}
\def\bra#1{\left\langle #1\right|}
\def\ket#1{\left| #1\right\rangle}
\def\abs#1{\left| #1\right|}
\def\vector#1{{\vec{#1}}}
\def\inv{^{\raise.15ex\hbox{${\scriptscriptstyle -}$}\kern-.05em 1}}
\def\pr#1{#1^\prime}  %prime
\def\lbar{{\lower.35ex\hbox{$\mathchar'26$}\mkern-10mu\lambda}} %lambda bar
\def\e#1{{\rm e}^{^{\textstyle#1}}}
\def\ee#1{\times 10^{#1} }
\def\imp{~\Rightarrow}
\def\coker{\mathop{\rm coker}}
\let\p=\partial
\let\<=\langle
\let\>=\rangle
\let\ad=\dagger
\let\txt=\textstyle
\let\h=\hbox
\let\+=\uparrow
\let\-=\downarrow
\def\dot{\!\cdot\!}
\def\vfilll{\vskip 0pt plus 1filll}
%
%%%%%%%%%%%%%%%%%%% end of intro %%%%%%%%%%%%%%%%%%%%%%%%%%%%%%%%%%%%%

\begin{titlepage}
\begin{flushright}
{IASSNS-HEP-97/128\\
CERN-TH/98-198\\
hep-ph/9806426\\
}
\end{flushright}
\vskip 2cm
\begin{center}
{\Large\bf The Fayet-Iliopoulos term in Type-I string theory
and M-theory}
\vskip 1cm
{\large
John March-Russell\footnote{Research
supported in part by U.S. Department of Energy contract
\#DE-FG02-90ER40542, and by the W.M.~Keck Foundation.
Alfred P. Sloan Foundation Fellow.}}\\
\vskip 0.5cm
{School of Natural Sciences\\
Institute for Advanced Study\\
Princeton, NJ~08540, USA\\}
\vskip 0.2cm
and\\
\vskip 0.2cm
{Theoretical Physics Division\footnote{On leave of absence from the
Institute for Advanced Study after June 2nd 1998.
Email: {\tt jmr@mail.cern.ch}}\\
CERN, CH-1211\\
Geneva 23, Switzerland}
\end{center}
\vskip .5cm
\begin{abstract}
The magnitude of the Fayet-Iliopoulos term is calculated for
compactifications of Type-I string theory and Horava-Witten
M-theory in which there exists a pseudo-anomalous U(1)$_X$.
Contrary to various conjectures, it is found that in leading order
in the perturbative expansion around the weakly-coupled M-theory
or Type-I limits, a result identical to that of the weakly-coupled
E$_8\times$E$_8$ heterotic string is obtained.  The result is
independent of the values chosen for the Type-I string scale
or the size of the M-theory 11th dimension, only depending upon
Newton's constant and the unified gauge coupling.  
\end{abstract}
\end{titlepage}
\setcounter{footnote}{0}
\setcounter{page}{1}
\setcounter{section}{0}
\setcounter{subsection}{0}
\setcounter{subsubsection}{0}

%%%%%%%%%%%%%%%%%%%%%%%%%%%%%%%%%%%%%%%%%%%%%%%%%%%%%%%%%%%%%%%%%%%%%%%

\section{Introduction} \label{sec:intro}

One of the most phenomenologically useful features of many 
string compactifications is the existence of a U(1) symmetry with apparent
field theoretic anomalies~\cite{DSW}.  These anomalies are cancelled by
a four-dimensional version of the Green-Schwarz mechanism~\cite{GS}, which
involves shifts of the model independent axion of string theory
under gauge and general coordinate transformations.  Such shifts are
able to compensate the field-theoretic anomalies of the pseudo-anomalous
U(1)$_X$ symmetry if and only if it has equal U(1)$_X$G$^2$ anomalies
for all gauge groups G as well as with gravity (here and in the following
we omit for simplicity the affine level factors $k_G$ which can easily
be reinstated if required).

As originally noticed by Dine, Seiberg, and Witten~\cite{DSW}, when this
mechanism is combined with supersymmetry, a Fayet-Iliopoulos (FI) term is
induced in the D-term for U(1)$_X$.  For the weakly-coupled
E$_8\times$E$_8$ heterotic string the magnitude of the induced
FI term depends on the field-theoretic anomaly coefficient,
the string coupling, $g_{\rm str}$, and
the reduced Planck mass $\mpl = M_{\rm Planck}/\sqrt{8\pi}$.  A
detailed calculation gives~\cite{DSW,atick,dine2}
\beq
\xi^2 = \frac{g_{\rm str}^2\Tr(Q_X)}{192\pi^2} \mpl^2, 
\label{eusual}
\eeq
where, following convention, this is expressed in terms of the
total U(1)$_X$(gravity)$^2$ anomaly proportional to $\Tr(Q_X)$.\footnote{
An interesting aspect of this result is that it evades the ``proof''
that a non-zero FI term is inconsistent
with local supersymmetry.}
 
This FI term allows many phenomenologically interesting applications.  These
include, an alternate non-grand-unified theory explanation of the successful
$\sin^2\th_w=3/8$ relation at the unification scale~\cite{ibanez},
the possible flavor universal communication of supersymmetry
breaking from a hidden sector to the minimal supersymmetric standard
model (MSSM)~\cite{anomalous}, and an explanation of the texture
of the fermion masses and mixings~\cite{IR}.  Many such 
applications require for their success the existence of
a hierarchy between the contributions
induced by the presence of the FI term and those contributions
arising from other supergravity or string theory modes.  Since the
strength of supergravity
corrections is controlled by the reduced Planck mass, the appropriate
expansion parameter is expected to be
\beq
\ep \equiv \frac{\xi^2}{\mpl^2} = \frac{g_{\rm str}^2\Tr(Q_X)}{192\pi^2}.
\label{eepsilon}
\eeq
For the value of $g_{\rm str}$ we may take the MSSM unified gauge
coupling, however, a remaining uncertainty in $\ep$ is the size
of the gravitational
anomaly coefficient.  For realistic string models, $\Tr(Q_X)$ is
certainly quite large; for instance, in one typical
semi-realistic example, (based on the free-fermionic
construction) $\Tr(Q_X) = 72/\sqrt{3}$~\cite{AP}.  However since
the gravitational anomaly is sensitive to all chiral states
in the model, both in the MSSM and hidden sectors, 
additional matter beyond the MSSM or large hidden sectors can increase 
$\Tr(Q_X)$ dramatically.  Thus
in the context of the weakly coupled E$_8\times$E$_8$ string
the expansion parameter is unlikely to be any smaller than
$\ep\sim 1/100$. 

The precise value of the expansion parameter $\ep$ is quite important
phenomenologically.  For example, in the case that the soft masses
of the MSSM squarks and sleptons are communicated by U(1)$_X$, the
MSSM gauginos, being necessarily uncharged under U(1)$_X$,
only receive soft masses via supergravity.  Thus the gaugino masses
are suppressed by the factor $\ep$ relative to the soft masses of the
MSSM squarks and sleptons (by assumption \ux\ charged -- at least for
the first and second generations)~\cite{anomalous}.  This leads to
either unacceptably light MSSM gauginos, or so heavy squarks and
sleptons that naturalness, or stability of the U(1)$_{\rm em}$ and
SU(3) color preserving minimum is threatened.
Although this problem is
ameliorated when a non-zero F-component for the dilaton superfield of
string theory is included~\cite{AHDM},
it is presently unclear if a fully consistent model is possible (especially
the question of whether $F$-terms for the other, non-universal, moduli are
induced).  In any case it is certainly interesting to ask how
the \ux\ scenarios become modified outside of the weakly-coupled
E$_8\times$E$_8$ case. 

Another of the proposed uses of the string-induced FI term is
D-term inflation~\cite{dinflation}.  In these models the inflationary
vacuum energy density is due to a non-zero D-term rather
than the more usual F-term contributions.
This has the advantage, over the F-term dominated models
of supersymmetric inflation of naturally possessing sufficiently flat inflaton
potentials for successful slow-roll inflation, even after the effects
of spontaneous supersymmetry breaking due to $V_{\rm vac}\neq 0$ are taken into
account.  
Since the magnitude of the FI term sets the size of the Hubble constant
during inflation, as well as its rate of change along the slow-roll
direction, it sets in turn the magnitude of the fluctuations in the
cosmic microwave background radiation (CMBR).  An analysis~\cite{dinflation}
shows that the observed amplitude of fluctuations in the CMBR requires
$\xi =6.6\times 10^{15}\gev$.  This is to be compared with the 
larger value, $\xi_{\rm string}\gsim 2\times 10^{17}\gev$, obtained
by substituting the deduced value of the MSSM unified coupling
into the weakly-coupled heterotic string theory prediction
Eq.(\ref{eusual}).
The most attractive proposal for the solution of this mismatch
is that the {\it weak-coupling} prediction for $\xi$ is modified
in the strong-coupling (Horava-Witten M-theory, or Type-I) limit.
This seems not unreasonable since it is well known that in such
a limit many E$_8\times$E$_8$ weak-coupling predictions are amended,
most notably the prediction for the scale of string
unification~\cite{wittenI}.

Given these various motivations for considering the FI term in more
general contexts, we show in the following sections
that a simple calculation of the induced FI
term in Type-I string theory and M-theory is possible.
The calculation is quite general, being independent of almost all
details of the compactification down to four dimensions.  However it
is important to keep in mind two assumptions under which we work: First,
we assume that the standard model gauge group arises from within
the perturbative excitations of the theory (rather than from D-brane
dynamics), and second, that there are not large corrections to the
normalisation of the kinetic terms of various fields arising
from corrections to the minimal Kahler potential.  Although such corrections
to the Kahler potential are certainly possible as we move away
from the weakly coupled M-theory or Type-I domain into the domain
of intermediate coupling, it seems very likely that the FI term itself 
is protected from such corrections by the surviving $N=1$ spacetime
supersymmetry.

\section{The FI term} \label{sec:FIterm}

The calculation of the magnitude of the FI term induced in a string
theory with an anomalous U(1)$_X$ symmetry is greatly simplified 
by a judicious use of supersymmetry together with the
anomaly constraints.  Since we are concerned with the
phenomenological applications of the FI-term, we are interested in
compactifications of either Type-I string theory or M-theory down
to four dimensions, in which the low-energy limit is an N=1 supersymmetric
field theory.  In this situation the dilaton $\phi$ and the
model-independent axion $a$ combine to form the lowest complex scalar
component of the dilaton chiral superfield,
$S = \exp(-2\phi) + i a + \ldots$.  Expanded out in terms of components,
the coupling of the dilaton superfield to a gauge field kinetic term of
a gauge group G reads
\beq
\frac{1}{4}\int d^2\th S W^{\al a}W_\al^a + {\rm h.c.}
= -\frac{1}{4e^{2\phi}} (F^a)^2 + \frac{1}{2e^{2\phi}} (D^a)^2 -
\frac{a}{4}F^a{\tilde F}^a +\ldots,
\label{eymke}
\eeq
where $\tilde F_{mn}^a = \frac{1}{2}\ep_{mnlp} F^{lpa}$.  Here
four-dimensional
Lorentz indices are denoted $m,n,\ldots$, and $a,b,\ldots$
are adjoint indices of G.  The expectation value of the dilaton
sets the value of the four-dimensional gauge couplings at an
appropriate scale, here always taken to be the string unification scale,
so $\< e^{2\phi}\> = g^2_{\rm unif}$.

It is easiest to calculate the FI term by starting with the
field-theoretic expression for 
the U(1)$_X$G$^2$ anomaly.  Such an anomaly implies that under a
U(1)$_X$ gauge transformation $v_m \to v_m + \p_m \om$ the
low-energy effective action is not invariant, but transforms as
\beq
\de \CL_{\rm eff} = \frac{\om A}{8\pi^2} {F\tilde F},
\label{elefftrans}
\eeq
where $A$ is the anomaly coefficient.  In a supersymmetric theory
$\om$ gets promoted to a chiral superfield
$\Om$, and the U(1)$_X$ vector superfield $V$ transforms
as $V\to V +i(\Om - \Om^*)/2$.  

The apparent anomaly in the low-energy effective action
Eq.(\ref{elefftrans}) is cancelled by assigning a shift in the 
axion under \ux\  gauge transformations.  In terms of
superfields,
\beq
\de S  = -i \frac{A}{8\pi^2} \Om,
\label{eashift}
\eeq
and then the shift in the gauge kinetic term Eq.(\ref{eymke}) compensates
the triangle anomaly Eq.(\ref{elefftrans}).  However, the non-trivial
transformation of $S$
implies that the conventional Kahler potential term for the 
dilaton superfield $\ln(S+S^*)$ is no longer invariant.  The
dilaton superfield kinetic energy term must be modified to
\beq
C^2 \int d^4\th \ln\left( S+S^* + \frac{A}{4\pi^2} V \right),
\label{eske}
\eeq
then the transformation of the \ux\  vector superfield cancels the 
variation.  (In Eq.(\ref{eske}) $C^2$ is an all important normalization
factor that will be calculated below from the underlying Type-I or
M-theory.)  The most important feature of Eq.(\ref{eske}) from our
point view is that when this modified kinetic term is expanded
out in component form, a term {\sl linear in the} \ux {\sl D-term}
occurs:
\beq
C^2 \left( \frac{1}{2}(\p \phi)^2 + \frac{e^{4\phi}}{4}(\p a)^2
+ \frac{Ae^{2\phi}}{16\pi^2} D + \ldots \right).
\label{eskecpt}
\eeq
This is nothing other than the FI term, $\xi^2$,
($\xi$ has mass dimension 1 in the conventions being followed here):
\beq
\xi^2 = {C^2 A \< e^{2\phi}\> \over 16\pi^2} =
{C^2 A g^2_{\rm unif} \over 16\pi^2}.
\label{exidef}
\eeq
Thus we see that supersymmetry allows us to calculate
the FI term given the coefficient of the anomaly, $A$, and the normalization,
$C^2$, of the model-independent axion kinetic energy, in the basis that the
axion itself is normalized so that 
it couples to the G gauge field strengths as in Eq.(\ref{eymke}). 

To fix $C$ it is necessary to consider the origin of the axion $a$
in string (or M-) theory.  Within the framework of an effective
10-dimensional string-derived supergravity, the model
independent axion arises from the 3-form field
strength $H_{\mu\nu\rho}$, which satisfies the
Green-Schwarz modified Bianchi identity
\beq
d H = -\tr F\wedge F + \tr R\wedge R .
\label{ebianchi}
\eeq
Once we compactify down to four dimensions we may define the effective
four-dimensional axion $a$ in terms of $H$ via
\beq
H_{mnp} = M^2 \ep_{mnpq} \p^q  a,
\label{ebara}
\eeq
where $M$ is a dimensionful normalization factor introduced so that
$a$ has mass dimension zero in four dimensions, and which
can be fixed by demanding that $a$ couples to $F\tilde{F}$ correctly.

The modified Bianchi identity, Eq.(\ref{ebianchi}), implies that
$a$ satisfies the equation of motion 
\beq
M^2 \p^2  a = \frac{1}{2} F^a {\tilde F}^a + (R{\rm -dependent~terms}),
\label{eabareom}
\eeq
which is equivalent to the statement that $a$ has the required
axion-like interaction term $aF{\tilde F}$ in the effective action
up to a normalization factor.
Furthermore, after compactification, the low-energy four-dimensional
kinetic energy for $H_{mnp}$
\beq
\CL_{\rm eff,4d} = \ldots + N H_{mnp} H^{mnp} +\ldots,
\label{eHke}
\eeq
induces the kinetic energy term for the axion.  To proceed we must convert
this to the normalization used in Eqs.(\ref{eymke}), (\ref{eskecpt}).  Fixing
the coupling of the axion to $F {\tilde F}$ gives the relation
$M^2 = 1/32N$, while we find that $C^2 = {e^{-4\phi}/32 N}$, and
thus using Eq.(\ref{exidef})
\beq 
\xi^2 = {A \over 512\pi^2 g^2_{\rm unif}N}.
\label{exi}
\eeq
Thus the calculation of the FI term is reduced to a computation
of the coefficient, $N$, of the $H$ kinetic term in the four-dimensional
low-energy effective action after compactification.

\section{The low-energy effective action in M-theory} \label{sec:Mtheory}

The Horava-Witten~\cite{HW} 11-dimensional low-energy supergravity action 
is, in component form, given by (see also~\cite{LOW}):
\bea
\CL_{\rm 11d} &=&\frac{1}{\ka^2} \int_{M} d^{11}x \sqrt{g} \left(
- \frac{1}{2} R  - \frac{1}{48} G_{IJKL}G^{IJKL} -
\frac{\sqrt{2}}{3456} \ep^{I_1\ldots I_{11}}
C_{I_1 I_2 I_3} G_{I_4 \ldots I_7}
G_{I_8\ldots I_{11}}  \right) \nonumber
\\
&+& \sum_{i=1,2} \frac{1}{8\pi(4\pi \ka^2)^{2/3}} \int_{\p M} d^{10}x
\left(- \tr F_i^2 + \frac{1}{2}\tr R^2 \right) + {\rm fermi~terms}.
\label{eHW}
\eea
This describes a system in which the supergravity modes propagate in
the 11-d bulk of $M$, while the E$^1_8\times$E$^2_8$ super
Yang-Mills (SYM) degrees of freedom are localized to the two 10-dimensional
orbifold hyperplanes $\p M_i$ $(i=1,2)$ of $M$ at $x^{11}=0,\pi\rho$
respectively.
The normalization of the gravitational part of the action is appropriate
for the 11th dimension being an $S^1$, of circumference $2\pi\rho$,
with the fields satisfying $Z_2$
orbifold conditions (the E$_8$ generators obey $\tr (T^a T^b)=\de^{ab}$).  
The 4-form field strength $G_{IJKL}$ satisfies a modified Bianchi
identity
\bea
(dG)_{11,IJKL}&=&-\frac{3\sqrt{2}}{2\pi}\left(\frac{\ka}{4\pi}\right)^{2/3}
\de(x^{11})\bigl( \tr F^1_{[IJ} F^1_{KL]} - \frac{1}{2} R_{[IJ}R_{KL]}
\bigr)\nonumber
\\
&-&\frac{3\sqrt{2}}{2\pi}\left(\frac{\ka}{4\pi}\right)^{2/3}
\de(x^{11}\, - \, \pi\rho)\bigl( \tr F^2_{[IJ} F^2_{KL]} -
\frac{1}{2} R_{[IJ}R_{KL]}
\bigr),
\label{eGbianchi}
\eea
and will provide, after compactification, the four-dimensional
model-independent axion $a$. 

Before we calculate the normalization of the axion kinetic term arising from
this action, it is necessary to eliminate $\ka$ and $\rho$ in terms of the 
4d gauge and gravitational couplings.  We can compactify the theory down to
four dimensions, preserving $N=1$ supersymmetry, by taking 
$M=S^1 \times R^4 \times K$, where $K$ is a (possibly deformed)
Calabi-Yau manifold, and where,
as usual, the spin connection is embedded in the E$^1_8$ gauge connection.
This leaves an E$_6$ SYM theory which can be further broken by Wilson lines, etc.  
(We assume in the following analysis that the visible MSSM gauge
group arises from the $i=1$ boundary theory.)  Denote the volume of the
Calabi-Yau measured at the $i=1$ boundary by $V(0)$.
As shown by Horava and Witten, in leading non-trivial
order in the expansion in $\ka^{2/3}$, the
volume of $K$ decreases linearly with $x^{11}$. The maximum allowed value
of $\rho$ is such that $V(x^{11}=\pi\rho)$ approximately equals the
Planck volume.  In any case,
compactifying the bulk supergravity action on $S^1 \times K$ leads to
\beq
\frac{1}{2\ka^2}\int d^{11}x\sqrt{g} R \to 
\frac{1}{2\ka^2}\int d^{4}x\sqrt{g^{(4)}}\frac{V(0)}{2} 2\pi\rho R^{(4)},
\label{emGcompact}
\eeq
where the factor of $V(0)/2$ arises from averaging over the Calabi-Yau
volume ($V(0)>>V(\pi\rho)$ in the scenarios where the hidden E$_8$ is
strongly coupled).  Comparing
this to the standard Einstein-Hilbert action in four dimensions
gives the usual result
\beq
G_N = {\ka^2 \over 8\pi^2 V(0)\rho}.
\label{eHWG}
\eeq
Similarly reducing the $i=1$ SYM boundary theory leads to an expression
for the unified gauge coupling $\al_{\rm unif}$ of the four-dimensional
theory
\beq
\al_{\rm unif} = {(4\pi \ka^2)^{2/3} \over 2V(0)}
\label{eHWal}
\eeq
It is important to recall that beyond leading order quantum corrections
modify these classical relations; however, the above expressions for
$G_N$ and $\al_{\rm unif}$ suffice to determine the FI term.

As discussed by Witten~\cite{wittenI}, the additional dependence
of $G_N$ on the size $\rho$ of the 11th dimension relative to $\al_{\rm unif}$ 
(as compared to the standard weak-coupling E$_8\times$E$_8$ result)
allows one to somewhat adjust the string prediction of the unification
scale.  Roughly speaking the four-dimensional MSSM unification scale can
be defined as the scale at which either string excitations, or the additional
six dimensions in which the gauge degrees of freedom propagate, first
become apparent.  An arbitrarily low unification scale is not achievable
within the Horava-Witten picture (at least if our world is situated at the
weakly coupled boundary), since the second E$_8$ SYM theory
moves into the strongly-coupled domain as $\rho$ increases.  Indeed
a calculation~\cite{wittenI} indicates that it is just possible
to take the length of the 11th dimension long enough so as to lower
the string unification scale to match that inferred in
the MSSM, $M_{\rm unif} \simeq 2\times 10^{16}\gev$.  In the following
we will keep the volume of the Calabi-Yau space and the length of the
11th dimension as free parameters.

Having completed this preliminary discussion, the effective 3-form
field strength $H$ is defined as
$G_{11,mnp} = \eta H_{mnp}$.  Here the normalization factor $\eta$ is necessary
so that the gauge dependent piece of the modified Bianchi identity for $H$
restricted to the $i=1$ boundary theory reads $dH = -\tr F^1\wedge F^1$, as in
Eq.(\ref{ebianchi}).  It follows from the M-theory version of the Bianchi identity
Eq.(\ref{eGbianchi}), that
\beq
\eta={3\sqrt{2} \over 4\pi^2\rho}\left({\ka\over 4\pi}\right)^{2/3}.
\label{eAnorm}
\eeq
The axion kinetic term arises from the $G^2$ term in Eq.(\ref{eHW}), which
expressed in terms of the correctly normalized 3-form field strength
reads, after compactification to four dimensions:
\beq
\frac{1}{12\ka^2}\int_M d^{11}x\sqrt{g} G^2_{ABC,11} \to
\frac{1}{12\ka^2}\pi\rho V(0)A^2\int_M d^{4}x\sqrt{g} H^2_{mnp}. 
\label{emake}
\eeq
Comparing to the general expression for the $H$ kinetic energy
leads gives $N$ which, using the relations Eqs.(\ref{eHWG}) and
(\ref{eHWal}) between $G_N$, $\al_{\rm unif}$ and $\ka,\rho$, and $V(0)$,
may be written
\bea
N &=& {3\over 32\pi^3 (4\pi)^{4/3}} {V(0)\over \rho\ka^{2/3}}\nonumber \\
  &=& {3G_N\over 16\pi \al^2_{\rm unif}}
\label{eNHWnorm} 
\eea
Finally, substituting this into the expression for the FI term, Eq.(\ref{exi}),
gives
\beq
\xi^2_{\rm HW} = {A g^2_{\rm unif} M^2_* \over 192\pi^2}.
\label{eHWxi}
\eeq
This is identical to the value obtained for the weakly-coupled
E$_8\times$E$_8$ heterotic string, {\it independent} of the size of the
11th dimension, or equivalently the unification scale in this
strongly-coupled E$_8\times$E$_8$ picture.

\section{Type-I string theory} \label{sec:typeI}

For Type-I string theory, the situation is quite similar.  We
start from the 10-dimensional supergravity Lagrangian derived
from its' low-energy limit~\cite{typeI}:
\bea
\CL_{\rm Type-I} =  &-& \int d^{10}x\sqrt{g}\biggl( {8\pi^7 e^{-2\phi_I}
\over (\al'_I)^4 }R + { e^{-\phi_I} \over 16 (\al'_I)^3 }F^a F^a\nonumber
\\
&+& {3 \over 2048 \pi^7 (\al'_I)^2} H^2 +\ldots \biggr).
\label{eIsugra}
\eea
Here $H$ is the 3-form field strength, $\phi_I$ the 10d dilaton of
Type-I string theory, and, as usual, $\al'_I$ has dimension of an 
inverse mass-squared.  If we compactify this theory on a Calabi-Yau
of volume $V$, the classical expressions for Newton's constant
and the 4d unified gauge coupling that follow from comparing to
the standard 4d action are
\bea
G_N &=& { e^{2\phi_I}(\al'_I)^4 \over 128 \pi^8 V}, \nonumber
\\
\al_{\rm unif} &=& { e^{\phi_I}(\al'_I)^3 \over  \pi V}.
\label{eIrelns}
\eea
As recently emphasised by Witten~\cite{wittenI}, it is now the different
dilaton dependencies of the gauge and gravitational pieces of the
action that allow one to adjust the string prediction of the unification
scale to match the inferred MSSM unification scale.  However unlike the
case in either the weakly-coupled E$_8 \times$E$_8$ heterotic
string or the standard Horava-Witten M-theory with E$^2_8$ strongly
coupled, the Type-I unification scale can be made arbitrarily low.  
In the Type-I case of interest, the unification scale
corresponds to leading order to $M_{\rm unif} \simeq 2\pi(V^{-1/6})$,
the mass of the first Kaluza-Klein excitations, and as this is
lowered, the type-I string coupling is also reduced, moving the theory
farther into the weakly coupled domain.  (However one must be careful that
the string worldsheet coupling is not becoming strong,
see Ref.~\cite{kaplunovsky} for an extensive discussion of these
issues.)

Independent of $V$, the expressions Eq.(\ref{eIrelns}) satisfy the
standard Type-I relation between the reduced Planck mass, $\mpl$,
and the gauge coupling,
$\al_{\rm unif}\mpl^2 = 128\pi^7 e^{-\phi_I}/\al'_I$.  (Note that
this relation differs by numerical factors from that stated in
Ref.~\cite{wittenI}.)  Solving for $\al'_I$ from Eqs.(\ref{eIrelns})
gives
\beq
(\al'_I)^2 = {\al_{\rm unif}^2 \mpl^2 V \over 16\pi^5}.
\label{eIalprime}
\eeq

As before, the quantity of interest for the calculation of the FI
term is the coefficient of the compactified 3-form field strength
kinetic term expressed in terms of $\al_{\rm unif}$, $M_{\rm unif}$
and $\mpl$.  Using Eq.(\ref{eIalprime}) leads to
\beq
N = {3 V \over 2048 \pi^7 (\al'_I)^2} =
{3 \over 128\pi^2 \al_{\rm unif}^2 \mpl^2},
\label{eIN}
\eeq
and thus from Eq.(\ref{exi}) we find that the
FI term in Type-I string theory is given by
\beq
\xi^2_I = {A M^2_{*} g^2_{\rm unif} \over 192\pi^2},
\label{eIxi}
\eeq
again identical to the result in the weakly-coupled 
E$_8\times$E$_8$ heterotic string.  This is independent of
the Type-I string scale, or equivalently the effective four-dimensional
unification scale, at least as long as one stays in the domain of
applicability of the above calculation.

\section{Comments}\label{sec:comments}

We have shown that a quite simple calculation of the induced
Fayet-Iliopoulos term is possible in the contexts of Type-I string
theory and Horava-Witten M-theory.
For fixed Newton's constant and four-dimensional unified gauge coupling 
the value of the FI term is independent of all details of the
underlying Type-I string theory or M-theory.  This includes the moduli
of the compactification down to four dimensions, the underlying string
coupling, and in the case of M-theory the length of the 11th dimension. 

One significant caveat to this work is that we have always assumed that
the gauge degrees of freedom of the MSSM appear in the perturbative
spectrum of the Type-I string theory or Horava-Witten M-theory.  Clearly
this is not a necessity, for the MSSM degrees of freedom could in
principle arise from D-brane excitations.  Interestingly, FI terms
for theories containing anomalous U(1)'s do arise in the
D-brane construction of six- and four-dimensional gauge theories (see
for example the discussions of Refs.~\cite{DM} and \cite{LPT}).  It would
certainly be worthwhile to consider such FI terms in the case of
realistic (or semi-realistic) D-brane models of the unified MSSM
interactions.

\section*{Acknowledgements}
I wish to thank K.~Dienes, C.~Kolda, and N.~Seiberg
for discussions, and M.~Dine for encouraging me to write up these
results.  I also wish to thank P.~Binetruy and P.~Ramond for discussions
of their closely related work prior to publication.

\end{document}